\documentclass[twocolumn,preprintnumbers,superscriptaddress,aps,prl]{revtex4-1}
\usepackage{amsmath}
\usepackage{amssymb}
\usepackage{graphicx}
\usepackage{hyperref}
\usepackage{xspace,slashed}
\usepackage[utf8]{inputenc}

\usepackage{booktabs}
\AtBeginDocument{
\heavyrulewidth=.08em
\lightrulewidth=.05em
\cmidrulewidth=.03em
\belowrulesep=.65ex
\belowbottomsep=0pt
\aboverulesep=.4ex
\abovetopsep=0pt
\cmidrulesep=\doublerulesep
\cmidrulekern=.5em
\defaultaddspace=.5em
}

\newcommand\xbj{x_{\scriptscriptstyle\rm Bj}}

\newcommand{\as}{\alpha_s}
\newcommand{\MSbar}{\overline{\text{MS}}}
\newcommand{\fb}{\;\mathrm{fb}}

\newcommand{\GeV}{\;\mathrm{GeV}}
\newcommand{\TeV}{\;\mathrm{TeV}}

\newcommand{\kslash}{/\!\!\!k}

\usepackage{xcolor}
\definecolor{light-gray}{gray}{0.8}

\begin{document}

\title{
How bright is the proton? \\
A precise determination of the photon parton distribution function}

\preprint{CERN-TH/2016-155}

\newcommand{\CERNaff}{CERN, Theoretical Physics Department, CH-1211
  Geneva 23, Switzerland} \newcommand{\UCSDaff}{Department of Physics,
  University of California at San Diego, La Jolla, CA 92093, USA}
\newcommand{\CNRSaff}{CNRS, UMR 7589, LPTHE, F-75005, Paris, France}
\newcommand{\MIBaff}{INFN, Sezione di Milano Bicocca, 20126 Milan,
  Italy} \newcommand{\OXFaff}{Rudolf Peierls Centre for Theoretical
  Physics, 1 Keble Road, University of Oxford, UK}

\author{Aneesh Manohar}
\affiliation{\UCSDaff}
\affiliation{\CERNaff}
\author{Paolo Nason}
\affiliation{\MIBaff}
\author{Gavin P. Salam}
\altaffiliation{On leave from \CNRSaff}
\affiliation{\CERNaff}
\author{Giulia Zanderighi}
\affiliation{\CERNaff}
\affiliation{\OXFaff}

\begin{abstract}
  It has become apparent in recent years that it is important, notably
  for a range of physics studies at the Large Hadron Collider, to have
  accurate knowledge on the distribution of photons in the proton.
  We show how the photon parton distribution function (PDF) can be
  determined in a model-independent manner, using electron--proton
  ($ep$) scattering data, in effect viewing the $ep\to e+X$ process as
  an electron scattering off the photon field of the proton.
  To this end, we consider an imaginary, beyond the Standard Model
  process with a flavour 
  changing photon--lepton vertex. We write its cross section in two
  ways, one in terms of proton structure functions, the other in terms
  of a photon distribution. Requiring their equivalence yields the
  photon distribution as an integral over proton structure functions.
  As a result of the good precision of $ep$ data, we constrain the
  photon PDF at the level of $1{-}2\%$ over a wide range of momentum
  fractions.
\end{abstract}


\maketitle

A fast-moving particle generates an associated electromagnetic field
which can be interpreted as a distribution of photons, as originally
calculated by Fermi, Weizs\"acker and
Williams~\cite{Fermi:1924tc,vonWeizsacker:1934nji,Williams:1934ad} for
point-like charges.
The corresponding determination of the photon distribution for
hadrons, specifically $f_{\gamma/p}$ for the proton, has however been
the subject of debate over recent years.

The photon distribution is small compared to that of the quarks and
gluons, since it is suppressed by a power of the electromagnetic
coupling $\alpha$.
Nevertheless, it has been realised in the past few years that its poor
knowledge is becoming a limiting factor in our ability to predict key
scattering reactions at CERN's Large Hadron Collider (LHC).
Notable examples are the production of the Higgs boson through $W/Z$
fusion~\cite{Ciccolini:2007ec}, or in association with an outgoing
weak boson~\cite{Denner:2011id}. For $W^\pm H$ production it is the
largest source of uncertainty~\cite{deFlorian:2016spz}.
The photon distribution is also potentially relevant for the
production of
lepton-pairs~\cite{Aad:2015auj,Aad:2016zzw,Accomando:2016tah,Alioli:2016fum,Bourilkov:2016qum},
top-quarks~\cite{Pagani:2016caq}, pairs of weak
bosons~\cite{Luszczak:2014mta,Denner:2015fca,Ababekri:2016kkj,Biedermann:2016yvs,Biedermann:2016guo,Yong:2016njr}
and generally enters into electroweak corrections for almost any LHC
process.
The diphoton excess around 750~GeV seen by ATLAS and
CMS~\cite{ATLAS-CONF-2015-081,CMS-PAS-EXO-15-004} has also generated
interest in understanding $f_{\gamma/p}$.

The two most widely used estimates of $f_{\gamma/p}$ are those
included in the \texttt{MRST2004QED}~\cite{Martin:2004dh} and
\texttt{NNPDF23QED}~\cite{Ball:2013hta} parametrisations of the proton
structure.
In the \texttt{NNPDF} approach, the photon distribution is constrained
mainly by LHC data on the production of pairs of leptons, $pp \to
\ell^+\ell^-$.
This is dominated by $q\bar q \to \ell^+\ell^-$, with a small
component from $\gamma\gamma \to \ell^+\ell^-$.
The drawback of this approach is that even with very small
uncertainties in $\ell^+\ell^-$ production data~\cite{Aad:2016zzw}, in
the QCD corrections to $q\bar q \to \ell^+\ell^-$ and in the quark and
anti-quark distributions, it is difficult to obtain high-precision
constraints on $f_{\gamma/p}$.

In the \texttt{MRST2004QED} approach, the photon is instead modeled.
It is assumed to be generated as emissions from free, point-like
quarks, using quark distributions fitted from deep-inelastic
scattering (DIS) and other data.
The free parameter in the model is an effective mass-scale below which
quarks stop radiating, which was taken in the range between
current-quark masses (a few MeV) and constituent-quark masses (a few
hundred MeV).
A more sophisticated approach~\cite{Gluck:2002fi} supplements a model
of the photon component generated from quarks (``inelastic'' part)
with a calculation of the ``elastic'' component (whose importance has
been understood at least since the early 1970's~\cite{Budnev:1974de})
generated by coherent radiation from the proton as a whole.
This was recently revived in
Refs.~\cite{Martin:2014nqa,Harland-Lang:2016apc,Harland-Lang:2016qjy}.
Such an approach was also adopted for the
\texttt{CT14qed\_inc}~\cite{Schmidt:2015zda} set, which further
constrains the effective mass scale in the inelastic component using
$ep \to e\gamma + X$ data~\cite{Chekanov:2009dq}, sensitive to the
photon in a limited momentum range through the reaction
$e\gamma \to e\gamma$~\cite{DeRujula:1998yq}.

In this article we point out that electron-proton ($ep$) scattering
data already contains all the information that is needed to accurately
determine $f_{\gamma/p}$.
It is common to think of $ep$ scattering as a process in which a
photon emitted from the electron probes the structure of the proton.
However one can equivalently think of it as an electron probing the
photon field generated by the proton itself. Thus the $ep$ scattering
cross section is necessarily connected with $f_{\gamma/p}$.
(This point of view is implicit also in 
Refs.~\cite{Anlauf:1991wr,Mukherjee:2003yh,Luszczak:2015aoa}.)
A simple way to make the connection manifest is to consider, instead
of $ep$ scattering, the fictitious process $l +p \to L +X$, where $l$
and $L$ are neutral leptons, with $l$ massless and $L$ massive with
mass $M$.
We assume a transition magnetic moment coupling of the form
$\mathcal{L}_{\text{int}} = (e/\Lambda) \overline L\, \sigma^{\mu \nu}
F_{\mu \nu}\, l$.
Here $e^2(\mu^2)/(4\pi) \equiv \alpha(\mu^2) $ is the $\MSbar$ QED coupling
evaluated at the scale $\mu$, and the arbitrary
scale $\Lambda \gg \sqrt s $ (where $\sqrt s$ is the centre-of-mass
energy) is introduced to ensure the correct dimensions.

The crucial observation that we rely on is inspired in part by Drees
and Zeppenfeld's study of supersymmetric particle production at $ep$
colliders~\cite{Drees:1988pp}: there are two ways of writing the
heavy-lepton production cross section $\sigma$, one in terms of
standard proton structure functions, $F_2$ and $F_L$ (or $F_1$), the other in
terms of the proton PDFs $f_{a/p}$, where the dominant flavour that
contributes will be $a = \gamma$.
Equating the latter result with the former will allow us to determine
$f_{\gamma/p}$.

We start with the inclusive cross section for $l(k) + p(p) \to
L(k^\prime) + X$.  Defining $q=k-k^\prime$, $Q^2=-q^2$ and
$\xbj=Q^2/(2 p q)$, we have
\begin{multline}
  \label{eq:crosssecform}
  \sigma = \frac{1}{4 p \cdot k} 
  \int \frac{d^4 q}{(2 \pi)^4 q^4} e_{\rm ph}^2(q^2) \left[ 4 \pi W_{\mu \nu}(p,q)
    \,L^{\mu \nu} (k, q) \right]
   \\ \times 2\pi\delta((k-q)^2-M^2)\,, 
\end{multline}
where the proton hadronic tensor (as defined in~\cite{Agashe:2014kda})
is given by $W_{\mu \nu}(p,q) = -g_{\mu\nu} F_1(\xbj,Q^2)+p_\mu
p_\nu/(pq) F_2(\xbj,Q^2)$ up to terms proportional to $q_{\mu}$,
$q_\nu$, and the leptonic tensor is $L^{\mu \nu}(k, q) = \frac12
(e^2_{\rm ph}(q^2)/\Lambda^2)\text{Tr}\left(\slashed{k}
  \left[\slashed{q},\gamma^\mu\right] (\kslash'+M)
  \left[\gamma^\nu,\slashed{q}\right]\right)$.
In Eq.~(\ref{eq:crosssecform}) we introduced the physical QED coupling
\begin{equation}
  e^2_{\rm ph}(q^2) = e^2(\mu^2)/(1-\Pi(q^2,\mu^2,e^2(\mu^2))),
\end{equation}
where $\Pi$ is the photon self energy and $\mu$ is the renormalisation
scale.
We stress that Eq.~(\ref{eq:crosssecform}) is accurate up to
corrections of order $\sqrt{s}/\Lambda$, since neither the
electromagnetic current nor the $ {\bar L} \gamma l$ vertex are
renormalised.

We find
\begin{multline}
  \label{eq:sigma-HL-simple}
 \hspace{-0.25cm} \sigma = \frac{c_0}{2\pi} 
  \int_x^{1-\frac{2xm_p}{M}} 
  \frac{dz}{z}
  \int^{Q_\text{max}^2}_{Q_\text{min}^2} 
  \frac{dQ^2}{Q^2} \alpha_{\rm ph}^2(-Q^2)   \Bigg[
  \biggl(
    2 
    -2 z 
    + z ^2\\
    + \frac{2 x ^2 m_p^2}{Q^2}
        +\frac{z ^2 Q^2}{M^2}
   -\frac{2 z  Q^2}{M^2}
    -\frac{2 x ^2 Q^2 m_p^2}{M^4}
  \biggr) F_2(x/z,Q^2) \\
  +\left(
    -z ^2
    -\frac{z ^2 Q^2}{2 M^2}
    +\frac{z ^2 Q^4}{2 M^4}
  \right)F_L(x/z,Q^2)
  \Bigg]\,,
\end{multline}
where  $x = M^2/(s-m_p^2)$, $m_p$ is the proton mass,
$F_L(x,Q^2)=(1+4m_p^2x^2/Q^2)F_2(x,Q^2)-2x F_1(x,Q^2)$
and $c_0=16\pi^2 /\Lambda^2$.
Assuming that $M^2  \gg m_p^2$, we have  $Q_{\min}^2 = x^2 m_p^2/(1-z)$ and
$Q_{\max}^2 = M^2(1-z)/z$.

\begin{figure}
  \centering
  \includegraphics[width=\columnwidth]{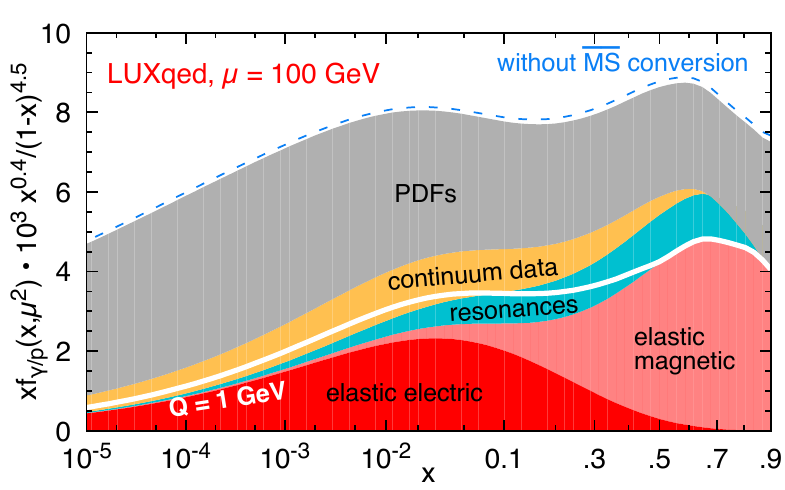}
  \caption{Contributions to the photon PDF at $\mu=100$~GeV,
    multiplied by $10^3 x^{0.4}/(1-x)^{4.5}$, from the various
    components discussed in the text. 
    The white line is the sum of the inelastic contribution from $Q^2 \le 1\,
    \text{(GeV)}^2$ in Eq.~(\ref{eq:xfgamma-MSbar}) and the full
    elastic contribution. 
    The result without the $\MSbar$ conversion term, i.e.\ the last
    term in Eq.~(\ref{eq:xfgamma-MSbar}), is given by the
    dashed blue line.
}
  \label{fig:breakup}
\end{figure}

The same result in terms of parton distributions can be written as
\begin{equation}
  \label{eq:coll-fact-basic}
  \sigma = c_0 \sum_{a} \int_x^1 \frac{dz}{z}\, 
   \hat\sigma_{a}(z,\mu^2)\, 
  \frac{M^2}{zs} f_{a/p}\left(\frac{M^2}{zs},\mu^2\right)\,,
\end{equation}
where in the $\MSbar$ factorisation scheme
\begin{multline}
  \label{eq:coll-sub-L-qe-alt}
  \hat \sigma_{a}(z,\mu^2) 
  = \alpha(\mu^2)\delta(1-z)\delta_{a\gamma} + \frac{\alpha^2(\mu^2)}{2\pi}\Bigg[ 
    -2 + 3z \, +
    \\
    + z p_{\gamma q}(z) \ln\frac{M^2(1-z)^2}{z\mu^2}
  \Bigg] \sum_{i\in \{q,\bar{q}\}} e_i^2 \delta_{a i}
  + \ldots\,,
\end{multline}
where $e_i$ is the charge of quark flavour $i$ and $z p_{\gamma q}(z)
= 1 + (1-z)^2$.
To understand which terms we choose to keep, observe that the photon
will be suppressed by $\alpha L$ relative to the quark and gluon
distributions, which are of order $(\as L)^n$, where $L=\ln\mu^2/m_p^2
\sim 1/\as$.
The contribution proportional to $F_2$ in
Eq.~(\ref{eq:sigma-HL-simple}) is of order $\alpha^2 
L (\as L)^n$, while that proportional to $F_L$ is of order $\alpha^2 (\as L)^n$.
We neglect terms that would be of order $\alpha^3L(\as L)^n$ or
$\alpha^2\as(\as L)^n$.
By requiring the equivalence of Eqs.~(\ref{eq:sigma-HL-simple}) and
(\ref{eq:coll-fact-basic}) up to the orders considered, one obtains
(in the $\MSbar$ scheme):
\begin{multline}
  \label{eq:xfgamma-MSbar}
  x f_{\gamma/p}(x,\mu^2) = 
  \frac{1}{2\pi \alpha(\mu^2)} \!
  \int_x^1
  \frac{dz}{z}
  \Bigg\{
  \int^{\frac{\mu^2}{1-z}}_{\frac{x^2 m_p^2}{1-z}} 
  \frac{dQ^2}{Q^2} \alpha^2(Q^2)
  \\
  \Bigg[\!
  \left(
    zp_{\gamma q}(z)
    + \frac{2 x ^2 m_p^2}{Q^2}
  \right)\! F_2(x/z,Q^2)
    -z ^2
  F_L\!\left(\frac{x}{z},Q^2\right)
  \Bigg]
  \\
  - \alpha^2(\mu^2) z^2 F_2\left(\frac{x}{z},\mu^2\right)
  \Bigg\}\,,
\end{multline}
where the result includes all terms of order $\alpha L\, (\as L)^{n}$,
$\alpha\, (\as L)^n$ and $\alpha^2 L^2\, (\as L)^{n}$
\footnote{The part of the integrand in square brackets appears also in
  Refs.~\cite{Anlauf:1991wr,Mukherjee:2003yh}, though the integration
  limits differ. Ref.~\cite{Luszczak:2015aoa} also obtained a result in
  terms of structure functions, however its Eq.~(3.25) is incompatible
  with leading-order QED DGLAP evolution.}.
Within our accuracy $\alpha_{\rm ph}(-Q^2)\approx \alpha(Q^2)$.
The conversion to the $\MSbar$ factorisation scheme, the last term in
Eq.~(\ref{eq:xfgamma-MSbar}), is small (see Fig.~\ref{fig:breakup}).

From Eq.~(\ref{eq:xfgamma-MSbar}) we have derived expressions up to order
$\alpha \as$ for the $P_{\gamma q}$, $P_{\gamma g}$ and $P_{\gamma
  \gamma}$ splitting functions using known results for the $F_2$ and
$F_L$ coefficient functions and for the QED $\beta$-function.
Those expressions agree with the results of a direct evaluation in
Ref.~\cite{deFlorian:2015ujt}.

The evaluation of Eq.~(\ref{eq:xfgamma-MSbar}) requires information on
$F_2$ and $F_L$.
Firstly (and somewhat unusually in the context of modern PDF fits), we will need the
elastic contributions to $F_2$ and $F_L$,
\begin{subequations}
\label{eq:F2L-elastic}
\begin{align}
  F_2^\text{el}(x,Q^2) &= \frac{[G_E(Q^2)]^2 + [G_M(Q^2)]^2 \tau}{1+\tau} \delta(1-x)\,,
  \\
  F_L^\text{el}(x,Q^2) &= \frac{[G_E(Q^2)]^2}{\tau} \delta(1-x)\,,
\end{align}
\end{subequations}
where $\tau = {Q^2}/{(4m_p^2)}$ and $G_E$ and $G_M$ are the electric
and magnetic Sachs form factors of the proton (see e.g.\ Eqs.(19) and
(20) of Ref.~\cite{Ricco:1998yr}).
A widely used approximation for $G_{E,M}$ is the dipole form
$G_E(Q^2)=1/(1+Q^2/m_\text{dip}^2)^{2}$, $G_M(Q^2)=\mu_pG_E(Q^2)$ with
$m_\text{dip}^2=0.71\GeV^2$ and $\mu_p\simeq 2.793$.
This form is of interest for understanding qualitative asymptotic
behaviours, predicting $f_{\gamma/p}(x) \sim \alpha (1-x)^4$ at large
$x$ dominated by the magnetic component, and $x f_{\gamma/p}(x) \sim
\alpha \ln 1/x$ at small $x$ dominated by the electric component.
However for accurate results, we will rather make use of a recent fit
to precise world data by the A1 collaboration~\cite{Bernauer:2013tpr},
which shows clear deviations from the dipole form, with an impact of
up to $10\%$ on the elastic part of $f_{\gamma/p}(x)$ for $x \lesssim
0.5$. The data constrains the form factors for $Q^2 \lesssim
10\GeV^2$.
At large $x$, Eq.~(\ref{eq:xfgamma-MSbar}) receives contributions only
from $Q^2 > x^2 m_p^2/(1-x)$, which implies that the elastic
contribution to $f_\gamma/p$ is known for $x \lesssim 0.9$.  Note that
the last term in Eq.~(\ref{eq:xfgamma-MSbar}) does not have an elastic
contribution for large $\mu^2$ because of the rapid drop-off of
$G_{E,M}$.

\begin{figure}
  \centering
  \includegraphics[width=\columnwidth]{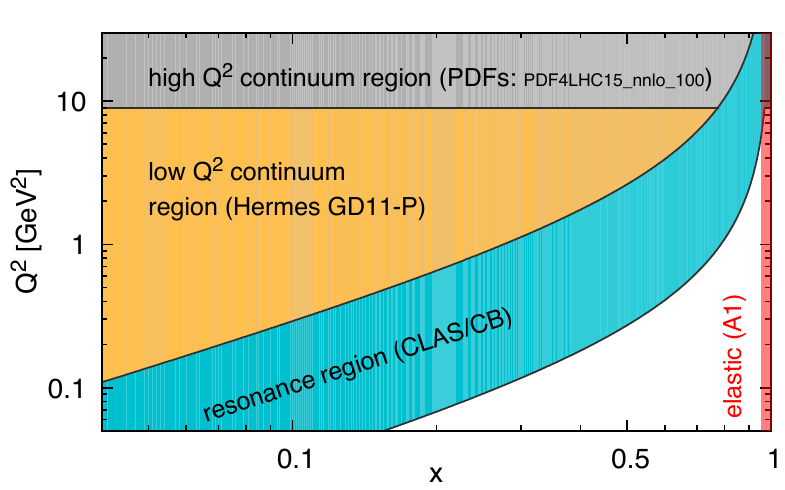}
  \caption{Our breakup of the $(x,Q^2)$ plane and the data
    for $F_{2}(x,Q^2)$ and $F_{L}(x,Q^2)$ we use in each region.
    The white region is inaccessible at leading order in QED.
  }
  \label{fig:kin-regions}
\end{figure}

The inelastic components of $F_2$ and $F_L$ contribute for $W^2 =
m_p^2 + Q^2 (1-x)/x > (m_p + m_{\pi^0})^2$.
One needs data over a large range of $x$ and $Q^2$. 
This is available thanks to a long history of $ep$ scattering studies.
We break the inelastic part of the $(x,Q^2)$ plane into three regions,
as illustrated in Fig.~\ref{fig:kin-regions}.
In the resonance region, $W^2 \lesssim 3.5\GeV^2$ we use a fit to data
by CLAS~\cite{Osipenko:2003bu}, and also consider an alternative fit
to the world data by Christy and Bosted (CB)~\cite{Christy:2007ve}.
In the low-$Q^2$ continuum region we use the GD11-P fit by
Hermes~\cite{Airapetian:2011nu} based on the ALLM parametric
form~\cite{Abramowicz:1991xz}.
Both the GD11-P and CB resonance fits are constrained by
photoproduction data, i.e.\ they extend down to $Q^2=0$.  The CLAS fit
also behaves sensibly there.
(Very low $Q^2$ values play little role because
the analytic properties of the $W^{\mu\nu}$ tensor
imply that $F_2$ vanishes as $Q^2$ at fixed $W^2$.)
These fits are for $F_2(x,Q^2)$. We also require $F_L$, or
equivalently $R=\sigma_L/\sigma_T$, which are related by
\begin{equation}
  \label{eq:FL-FN-R}
  F_L(x,Q^2) = 
  F_2(x,Q^2) \left(1 + \frac{4m_p^2 x^2}{Q^2} \right)
  \frac{R(x,Q^2)}{1 + R(x,Q^2)}\,,
\end{equation}
and we use the parametrisation for $R$ from
HERMES~\cite{Airapetian:2011nu}, extended to vanish smoothly as $Q^2
\to0$.
The leading twist contribution to $F_L$ is suppressed by
$\alpha_s(Q^2)/(4\pi)$.
At high $Q^2$ we determine $F_2$ and $F_L$ from the
\texttt{PDF4LHC15\_nnlo\_100}~\cite{Butterworth:2015oua} merger of
next-to-next-to-leading order (NNLO)~\cite{Moch:2004pa,Vogt:2004mw} global PDF
fits~\cite{Ball:2014uwa,Harland-Lang:2014zoa,Dulat:2015mca}, using
massless NNLO coefficient
functions~\cite{vanNeerven:1991nn,Zijlstra:1992qd,vanNeerven:1999ca,vanNeerven:2000uj}
implemented in \texttt{HOPPET}~\cite{Salam:2008qg,Cacciari:2015jma,Dreyer:2016oyx}.

In Fig.~\ref{fig:breakup}
we show the various contributions to our photon PDF, which we dub
``\texttt{LUXqed}'', as a function of $x$, for a representative scale
choice of $\mu=100\GeV$.
There is a sizeable elastic contribution, with an important magnetic
component at large values of $x$. The white line represents
contributions arising from the $Q^2<1$ region of all the structure
functions, including the full elastic contribution.  For the accuracy
we are aiming at, all contributions that we have considered, shown in
Fig.~\ref{fig:breakup}, have to be included, and inelastic
contributions with $Q^2<1$ cannot be neglected.
The photon momentum fraction is $0.43\%$ at $\mu=100 \GeV$.

\begin{figure}
  \centering
  \includegraphics[width=\columnwidth]{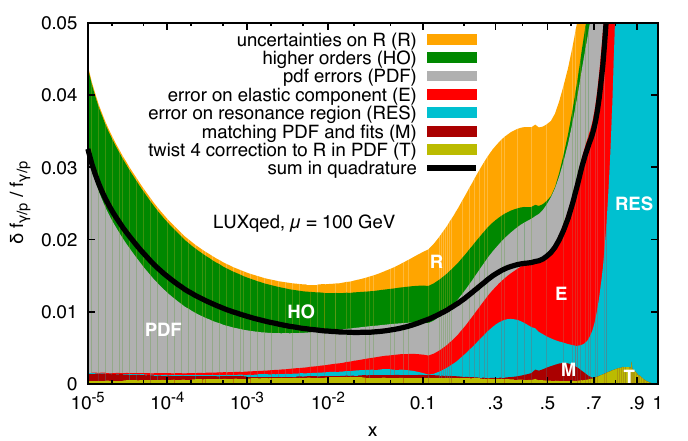}
  \caption{Linearly stacked relative uncertainties on the photon PDF,
    from all sources we have considered, and their total sum in
    quadrature shown as a black line, which is our final uncertainty.}
  \label{fig:uncertainties}
\end{figure}

In Fig.~\ref{fig:uncertainties}
we show the sources contributing to the uncertainty on our calculation
of $f_{\gamma/p}$ at our reference scale $\mu=100\GeV$.
They are stacked linearly and consist of:
a conservative estimate of $\pm50\%$ for the uncertainty on
$R=\sigma_L/\sigma_T$ at scales $Q^2<9\GeV^2$ (R);
standard $68\%$CL uncertainties on the PDFs, applied to scales $Q^2
\ge 9\GeV^2$ (PDF);
a conservative estimate of the uncertainty on the elastic form
factors, equal to the sum in quadrature of the fit error
and of the estimated size of the two-photon exchange
contribution in~\cite{Bernauer:2013tpr} (E);
an estimate of the uncertainty in the resonance region taken as the
difference between the CLAS and CB fits (RES);
a systematic uncertainty due to the choice of the transition scale
between the HERMES $F_2$ fit and the perturbative determination from
the PDFs, obtained by reducing the transition scale from $9$ to
$5\GeV^2$ (M);
missing higher order effects, estimated using a modification of
Eq.~\eqref{eq:xfgamma-MSbar}, with the upper bound of the $Q^2$
integration set to $\mu^2$ and the last term adjusted to maintain
$\alpha^2 (\as L)^n$ accuracy (HO);
a potential twist-4 contribution to $F_L$ parametrised as a factor
$(1+5.5\GeV^2/Q^2)$~\cite{Cooper-Sarkar:2016foi} for $Q^2 \ge 9\,{\rm
  GeV}^2$ (T).
One-sided errors are all symmetrised.
Our final uncertainty, shown as a solid line in
Fig.~\ref{fig:uncertainties}, is obtained by combining all sources in
quadrature and is about 1-2\% over a large range of $x$ values.

\begin{figure}
  \centering
  \includegraphics[width=\columnwidth]{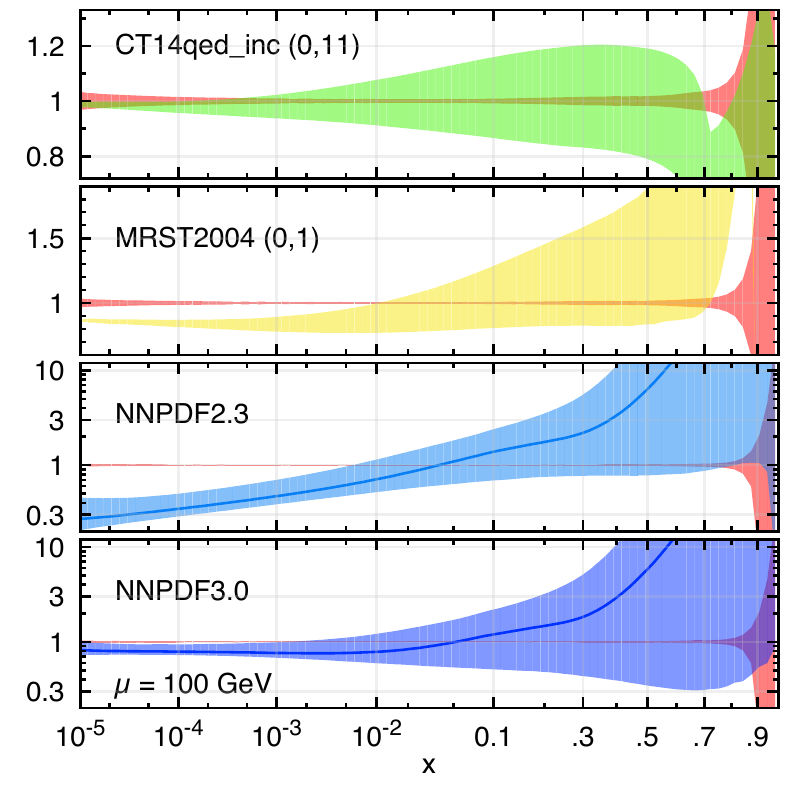}
  \caption{The ratio of common PDF sets to our \texttt{LUXqed} result, along
    with the \texttt{LUXqed} uncertainty band (light red).
    The \texttt{CT14} and \texttt{MRST} bands correspond to the range from the PDF members shown
    in brackets (68\% cl.\ in \texttt{CT14}'s case).
    The \texttt{NNPDF} bands span from $\max(\mu_r-\sigma_r,r_{16})$ to $\mu_r+\sigma_r$,
    where $\mu_r$ is the average (represented by the blue line), $\sigma_r$
    is the standard deviation over replicas, and $r_{16}$ denotes the
    16$^\text{th}$ percentile among replicas.
    Note the different $y$-axes for the panels.  }
  \label{fig:comparison}
\end{figure}

In Fig.~\ref{fig:comparison}
we compare our \texttt{LUXqed} result for the $\MSbar$ $f_{\gamma/p}$
to determinations available publicly within
\texttt{LHAPDF}~\cite{Buckley:2014ana}.
Of the model-based estimates, \texttt{CT14qed\_inc}~\cite{Schmidt:2015zda} and
\texttt{MRST2004}~\cite{Martin:2004dh}, \texttt{CT14qed\_inc}
is in good agreement with \texttt{LUXqed} within its uncertainties.
Its model for the inelastic component is constrained by $e p \to
e\gamma + X$ data from ZEUS~\cite{Chekanov:2009dq} and includes
an elastic component.  Note however that, for the neutron,
\texttt{CT14qed\_inc} neglects the important neutron magnetic form
factor.
As for the model-independent determinations,
\texttt{NNPDF30}~\cite{Bertone:2016ume}, which notably extends
\texttt{NNPDF23}~\cite{Ball:2013hta} with full treatment of $\alpha
(\as L)^n$ terms in the evolution~\cite{Bertone:2013vaa}, almost
agrees with our result at small $x$.
At large $x$ its band overlaps with our result, but the central value
and error are both much larger.

\begin{figure}
  \centering
  \includegraphics[width=\columnwidth]{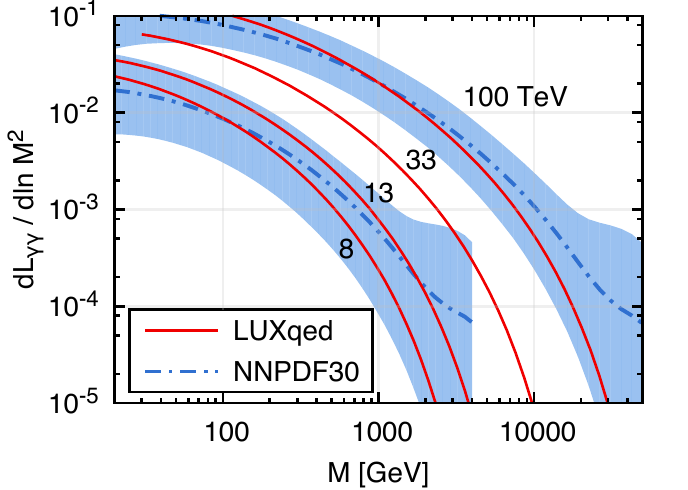}
  \caption{$\gamma\gamma$ luminosity in $pp$ collisions as a function of the
    $\gamma\gamma$ invariant
    mass $M$, at four collider centre-of-mass energies. 
    The \texttt{NNPDF30} results are shown only for $8$ and $100\TeV$.
    The uncertainty of our \texttt{LUXqed} results is smaller than the
    width of the lines.  }
  \label{fig:lumi}
\end{figure}

Similar features are visible in the corresponding $\gamma\gamma$
partonic luminosities, defined as
\begin{align}
  \frac{dL_{\gamma \gamma}}{d\ln M^2} =
  \frac{M^2}{s} \int \frac{dz}{z}\, f_{\gamma/p}(z,M^2)\, 
  f_{\gamma/p}\left( \frac{M^2}{z s}, M^2\right),
\end{align}
and shown in Fig.~\ref{fig:lumi}, as a function of the $\gamma\gamma$
invariant mass $M$, for several centre-of-mass energies.

As an application, we consider $pp \to HW^+(\to\ell^+\nu) + X$ at
$\sqrt{s}=13\TeV$, for which the total cross section without
photon-induced contributions is
$91.2\pm1.8\fb$~\cite{MelladoGarcia:2150771}, with the error dominated
by (non-photonic) PDF uncertainties.
Using \texttt{HAWK}~2.0.1~\cite{Denner:2014cla}, we find a photon-induced
contribution of $5.5^{+4.3}_{-2.9}\fb$ with \texttt{NNPDF30}, to be
compared to $4.4\pm0.1 \fb$ with \texttt{LUXqed}.

In conclusion, we have obtained a formula
(i.e. Eq.~(\ref{eq:xfgamma-MSbar})) for the $\MSbar$ photon PDF in
terms of the proton structure functions, which includes all terms of
order $\alpha L\, (\as L)^{n}$, $\alpha\, (\as L)^n$ and $\alpha^2
L^2\, (\as L)^{n}$. Our method can be easily generalised to higher
orders in $\as$ and holds for any hadronic bound state.
Using current experimental information on $F_{2}$ and $F_L$ for
protons we obtain a
photon PDF with much smaller uncertainties than existing
determinations, as can be seen from Fig.~\ref{fig:comparison}.
The photon PDF has a substantial contribution from the elastic form
factor ($\sim \! 20$\%) and from the resonance region ($\sim \! 5$\%)
even for high values of $\mu \sim 100{-}1000$~GeV.
Our photon distribution, incorporating quarks and gluons from
\texttt{PDF4LHC15\_nnlo\_100}~\cite{Butterworth:2015oua} and evolved
with a QED-extended version of \texttt{HOPPET} is available as part of
the \texttt{LHAPDF} library as the
\texttt{LUXqed\_PDF4LHC15\_nnlo\_100} set and from
\url{http://cern.ch/luxqed}.
Note that it is only valid for scales $\mu > 10 \GeV$.

More details of our analysis, including a derivation using PDF
operators, computation of splitting functions, higher order
corrections to Eq. (\ref{eq:xfgamma-MSbar}), as well as an extension
to the polarized case will be given in a longer
publication~\cite{MNSZ}.


\acknowledgments

We would like to thank Silvano Simula who provided us with a code for
the CLAS parametrisation, Jan Bernauer for discussions of the A1
results and fits and Gunar Schnell for bringing the HERMES
\mbox{GD11-P} fit to our attention and providing the corresponding
code.
We also thank Markus Diehl, Stefan Dittmaier, Stefano Forte, Kirill
Melnikov and Jesse Thaler for helpful discussions.
This work was supported in part by ERC Consolidator Grant HICCUP (No.\
614577), ERC
Advanced Grant Higgs@LHC (No.\ 321133), a grant from the Simons Foundation (\#340281
to Aneesh Manohar), by DOE grant DE-SC0009919, and NSF grant NSF
PHY11-25915.
We also acknowledge MITP (GZ) and KITP (GPS, GZ) for hospitality while
this work was being completed.

\bibliography{photon}

\end{document}